\ifdefined\pdfoutput
\pdfoutput=1
\fi
\documentclass[11pt]{article}

\usepackage[final]{acl}

\usepackage{times}
\usepackage{amsmath}
\usepackage{amsfonts}
\usepackage{latexsym}
\usepackage{booktabs}
\usepackage{cleveref}

\usepackage{xcolor}
\usepackage{multirow}
\usepackage{todonotes}
\usepackage{tikz}
\usepackage{tabularx}

\newcolumntype{Y}{>{\centering\arraybackslash\scriptsize}X}

\usepackage[svgnames]{xcolor}
\usepackage{colortbl}
\usepackage{makecell} 
\usepackage[T1]{fontenc}
\usepackage{soul} 

\usetikzlibrary{positioning,fit,calc,arrows.meta,shapes.misc}

\usepackage[most]{tcolorbox}

\definecolor{promptbg}{RGB}{248,249,251}
\definecolor{promptborder}{RGB}{209,213,219}
\definecolor{prompttitle}{RGB}{31,41,55}
\definecolor{promptrolebg}{RGB}{229,231,235}
\definecolor{promptrolesys}{RGB}{30,64,175}
\definecolor{promptroleusr}{RGB}{124,45,18}
\definecolor{promptroleasg}{RGB}{22,101,52}

\DeclareMathOperator*{\argmax}{arg\,max}

\usepackage[T1]{fontenc}

\usepackage[utf8]{inputenc}

\usepackage{microtype}
\usepackage{xspace}

\usepackage{inconsolata}

\usepackage{graphicx}
\usepackage{caption}
\graphicspath{{Media/}{Media/Credit_Assignments/}}

\newcommand{\authorcomment}[3]{{\textcolor{#2}{[#1: #3]}}}
\renewcommand{\authorcomment}[3]{}

\definecolor{orba}{RGB}{255, 143, 0}  
\definecolor{salmon}{RGB}{255, 100, 100}  

\newcommand{\frankenreport}{\textsc{\textbf{FrankenReport}}\xspace}

\newenvironment{llmjudgeprompt}[2][]{%
  \begin{tcolorbox}[
      breakable,
      enhanced,
      colback=promptbg,
      colframe=promptborder,
      coltitle=prompttitle,
      fonttitle=\bfseries,
      boxrule=0.4pt,
      arc=1.5mm,
      outer arc=1.5mm,
      left=2mm,right=2mm,top=2mm,bottom=2mm,
      boxsep=1.5mm,
      title={#2},
      attach boxed title to top left={yshift=-1mm,xshift=1mm},
      boxed title style={
        colback=promptbg,
        colframe=promptborder,
        boxrule=0.4pt
      },
      fontupper=\ttfamily\small,
      #1
    ]
  }{%
  \end{tcolorbox}
}

\newcommand{\promptmeta}[1]{%
  \vspace{1mm}%
  {\normalfont\footnotesize\itshape #1}\par
  \vspace{1.5mm}%
}

\newcommand{\rolebadge}[2]{%
  {\normalfont\scriptsize
    \colorbox{promptrolebg}{%
      \textcolor{#2}{\textsf{\textbf{#1}}}%
    }%
  }%
}

\newcommand{\userrole}{\rolebadge{USER}{promptroleusr}}

\definecolor{hallucinationbg}{HTML}{FFEBEB} 
\definecolor{hallucinationtxt}{HTML}{D81B60} 
\definecolor{factgainbg}{HTML}{E6FFEC} 
\definecolor{factgaintxt}{HTML}{1E88E5}
\definecolor{redundanttxt}{HTML}{888888} 

\newcommand{\bad}[1]{\textbf{\textit{\textcolor{hallucinationtxt}{#1}}}}
\newcommand{\good}[1]{\textbf{\textcolor{factgaintxt}{#1}}}
\newcommand{\meh}[1]{\textcolor{redundanttxt}{#1}}

\title{FrankenReport: Early Exiting in Long-Form Generation \\ Using Expected Value of Computation}

\author{
  \textbf{Zhengping Jiang\textsuperscript{$\spadesuit$}}\thanks{Work performed during an internship at Microsoft.} \quad
  \textbf{Gonzalo Ramos\textsuperscript{$\heartsuit$}} \quad
  \textbf{Jina Suh\textsuperscript{$\heartsuit$}} \\
  \textbf{Shiqian Rachel Ng\textsuperscript{$\heartsuit$}} \quad
  \textbf{Elias Stengel-Eskin\textsuperscript{$\heartsuit$}} \quad
  \textbf{Justin Svegliato\textsuperscript{$\heartsuit$}} \\
  \textbf{Benjamin Van Durme\textsuperscript{$\heartsuit$}} \quad
  \textbf{Andy Huntington\textsuperscript{$\heartsuit$}} \quad
  \textbf{Sam Thomson\textsuperscript{$\heartsuit$}} \\
  \textsuperscript{$\spadesuit$}Johns Hopkins University \quad
  \textsuperscript{$\heartsuit$}Microsoft \\
  \texttt{zjiang31@jh.edu}
}

\begin{document}
\maketitle
\begin{abstract}

  While deep research systems address interactive information-seeking needs impressively, their real-world deployments face latency and resource-consumption challenges.
  We present \frankenreport, an interface for long-form knowledge-seeking report generation that supports adaptive early exiting per section: it evaluates intermediate outputs during generation and predicts whether further targeted computation will yield significant quality gains.
  In a simulation study, \frankenreport outperforms random allocation baselines by a large margin (up to $4\times$) under low budgets and smoothly recovers full-pipeline quality as the budget grows, showing that future quality gains are predictable from intermediate drafts.
  Through experiments and user studies, we further show that despite varying preferences across users and topics, \frankenreport adapts to simple, natural user feedback as efficiently as methods requiring much costlier supervision such as generated drafts and explicit rationales.

\end{abstract}
\section{Introduction}
Deep research systems generate structured, multi-section long-form responses---often called reports---that resemble Wikipedia-style articles and 
include explicit citations \citep{shao2024assisting, qian2023webbrain, walden2025grounded}.
Recent systems such as OpenAI Deep Research, Gemini Deep Research, and DeerFlow illustrate the growing deployment of multi-step research agents in knowledge-intensive settings
\citep{openai2025deepresearch, google2025deepresearch, bytedance2025deerflow, fan2025understanding}.
These capabilities rely on iterative retrieval and reasoning pipelines that incur substantial computational and monetary
cost, and lead to latency for the user.

\begin{table}[!t]
  \small
  \renewcommand{\arraystretch}{1.25} 
  \setlength{\tabcolsep}{2pt} 

  \begin{tabularx}{\linewidth}{ @{}m{1.3cm} Y Y >{\centering\arraybackslash}m{1.0cm} }
    \toprule
    \textbf{Case} & \textbf{Low Effort} & \textbf{High Effort} & \textbf{Action} \\
    \midrule

    \makecell[l]{\scriptsize \textbf{Jimmy}\\\scriptsize \textbf{Zavala}} &
    {\textit{\textcolor{gray}{Insufficient information available regarding early life and career details...}}} &
    {Born \good{July 12, 1955}. Influenced by \good{Junior Walker} and \good{King Curtis}, shaping his style...} &
    \makecell[c]{\textcolor{DarkGreen}{\small$\uparrow$} {\scriptsize\textbf{Run}}\\[-3pt]{\tiny(Knowledge)}} \\

    \midrule 

    \makecell[l]{\scriptsize \textbf{Sahara}\\\scriptsize \textbf{Desert}} &
    {The Sahara Desert is the world’s largest hot desert, covering approx 9.2M sq km across North Africa...} &
    {The Sahara is the largest hot desert... \meh{spanning approx 3000 miles East to West across 11 countries...}} &
    \makecell[c]{\textcolor{orange}{\small$\odot$} {\scriptsize\textbf{Stop}}\\[-3pt]{\tiny(Marginal)}} \\

    \midrule

    \makecell[l]{\scriptsize \textbf{GlycoBlue}} &
    {GlycoBlue acts as a \bad{PCR additive to improve yield} and specificity of reactions...} &
    {It enhances the \bad{visibility of DNA bands during electrophoresis} and stabilizes the mixture...} &
    \makecell[c]{\textcolor{orange}{\small$\odot$} {\scriptsize\textbf{Stop}}\\[-3pt]{\tiny(Unhelpful)}} \\
    \bottomrule
  \end{tabularx}

  \caption{More computational effort does not guarantee better results, as illustrated in the ``low effort'' and ``high effort'' columns. \meh{Gray} texts add little value, \bad{Red} are factually wrong and \good{Blue} are additional accurate facts.
  \frankenreport predicts the quality gain from additional computational effort and takes actions accordingly, only expending effort on cases with a high estimated quality gain.}
  \label{tab:teaser}
\end{table}

High computational cost does not guarantee better user-perceived quality. In interactive systems, response latency itself shapes perceived quality and usage intentions \citep{gnewuch2022opposing}, while expressed confidence can influence users independently of substantive accuracy \citep{ma2024you,xu2025confronting,fernandes2025ai,li2025confidence,wang2025impact}. Consequently, additional compute frequently yields diminishing returns. Users prioritize speed, tolerating delays only for noticeable, justified improvements \citep{abbas2022understanding,zhang2024explaining,wang2025effects}.

This motivates a core control problem: determining when extra compute meaningfully improves perceived quality. While early exiting succeeds in Transformer inference \citep{xin2020deeb,xin2021berxit}, report generation differs from classification. Unlike binary correctness, the marginal value of generative compute is subjective and multi-dimensional, making a priori resource allocation difficult.

Our approach is motivated by a simple observation: deep research pipelines produce informative intermediate artifacts: outlines, gists, and partial drafts that already capture much of the final report's value. These outputs provide early signals of organization, coverage, and factual grounding. Building on this, we introduce \frankenreport, a progressive report generation interface that surfaces increasingly refined intermediate results, starting from a lightweight sketch and incrementally improving as execution proceeds, while a fine-tuned LLM with a regression head predicts the scale of the potential quality improvements to enable adaptive early exiting for more efficient generation. The name reflects how the final report is assembled by section from components that may have undergone different depths of refinement.

Our experiments show that even the earliest structural sketch provides strong predictive signals: our regression model can predict a quality gap on multiple dimensions much more effectively from seeing this sketch than from the query alone (\autoref{tab:predictor-effectiveness}), and achieve greater quality improvements under limited computation budget as compared to a random exiting baseline (\autoref{fig:credit-assignments}).
A 100-participant study confirms that user preferences are diverse but are learnable, demanding personalized computation budget allocation. 
We show that \frankenreport adapts to diverse preferences with sample efficiency comparable to methods requiring much costlier supervision (e.g., full drafts or rationales).

  \section{Related Work}

We target single-turn, deep research workflows spanning applications such as business analysis, market research, and clinical summarization \citep{xu2025comprehensive}. Deployed systems explicitly pair comprehensive reports with short turnaround times \citep{perplexity2025deepresearch}, motivating compute-aware generation strategies.

  \subsection{Early Exiting in Transformer Models}
  \label{sec:early-exit-related-work-anchor}
  Early exiting augments deep networks with intermediate classifiers that halt inference once confidence is sufficient, originating in vision models~\citep{teerapittayanon2016branchynet} and later adopted in Transformers. Analyses of decoder-only LLMs show that shallow layers already encode many correct predictions~\citep{nostalgebraist2020logitlens, belrose2023eliciting}, and methods such as DeeBERT, FastBERT, and PABEE demonstrate large latency reductions with minimal accuracy loss~\citep{xin2020deeb,zhu2020fastbert,zhou2020pabee}. Subsequent work extends these ideas to parameter-efficient tuning, adaptive routing, RL-based exit policies, and large-scale generation~\citep{xin2021berxit,li2023odebert,hu2023smartbert,chen2023pfberxit,li2024consistentee}.

  \subsection{Early Exiting from Reasoning Chains}
  As extended reasoning chains have become common in LLM applications and can be computationally expensive, recent work targets truncating the reasoning chain itself, instead of early exiting from the network. \citet{jurayj2025your} find that increasing test-time compute raises confidence on correct responses. Early-stopping approaches such as ES-CoT~\citep{mao2025early}, HALT-CoT~\citep{laaouach2025halt}, and LEASH~\citep{quamar2025logit} exploit intermediate-answer convergence, answer entropy, or token-level logits to decide when to stop. Other approaches include REFRAIN~\citep{sun2025stop}, which detects reflective but redundant reasoning and adapts its stopping threshold with a bandit controller, or using injected exit instructions or an extrinsic completion verifier to decide when to stop~\citep{lu2025runaway}.

  \subsection{Semantic Routing for LLM Cascades}
  Prior work routes queries among models of varying capability to balance cost and quality. FrugalGPT~\citep{chen2024frugalgpt} escalates to stronger models only when needed; learned routers predict difficulty or adjust routing online~\citep{ding2024hybrid,ding2025bestroute}; and other methods route by semantic intent~\citep{hari2023tryage} or apply multi-agent refinement selectively~\citep{chen-etal-2025-magicore}. Complementary approaches escalate when outputs disagree~\citep{yue2024large,kolawole2025abc,soiffer-etal-2025-semantic}, estimate uncertainty via semantic entropy or confidence tokens~\citep{mehradfar2025semanticentropy,li2025conftuner}, or combine routing with speculative computation~\citep{narasimhan2025speculative}. Similar principles now appear in deployed systems such as GPT-5~\citep{openai2025gpt5}.

  \subsection{Wikipedia-style Report Generation}
  We build on the STORM pipeline~\citep{shao2024assisting}. STORM extracts $K$ perspectives on a knowledge-seeking query and runs $K$ parallel \textit{dialogue threads} between a Questioner and a Topic Expert (both LLM agents); over $T$ turns the Questioner issues search queries and the Expert retrieves and summarizes documents. The passages are then aggregated into an outline and expanded into section-level prose.

\begin{figure}[!htpb]
    \includegraphics[width=\linewidth]{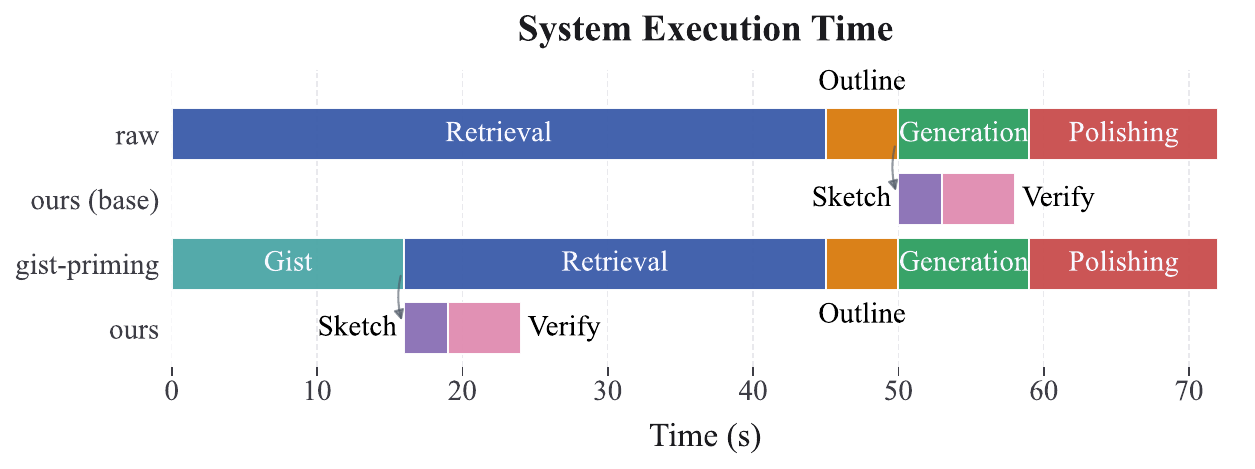}
    \caption{
    Mean wall-clock timing for a single request. The top pair shows the original STORM pipeline and its late Sketch/Verify checkpoints; the bottom pair shows gist-primed planning and its earlier checkpoints (\autoref{sec:method}). Within each pair, the thin checkpoint row shares the timeline of the pipeline row above it. Sections execute in parallel after the shared planning work.}
    \label{fig:gantt-plot}
  \end{figure}

  We formulate \emph{budgeted early exiting} for multi-stage, tool-augmented pipelines such as STORM, jointly deciding \emph{when} and \emph{where} to spend compute across sections, refinement operators, and tool use. Building on prior work, our work contributes: \textbf{(1)~Problem formulation:} constrained resource allocation over sections and refinements instead of per-token or per-layer stopping; \textbf{(2)~Method:} a future-utility predictor that estimates per-section quality gains from further refinement and guides selective computation and early termination (\autoref{fig:gantt-plot}); and \textbf{(3)~Evaluation:} quality-cost trade-off curves that characterize budget-allocation performance, especially in low-budget regimes, across user-relevant quality dimensions.

  \begin{figure*}[!htbp]
    \includegraphics[width=\textwidth, trim={.2 .2 .2 0.2cm}, clip]{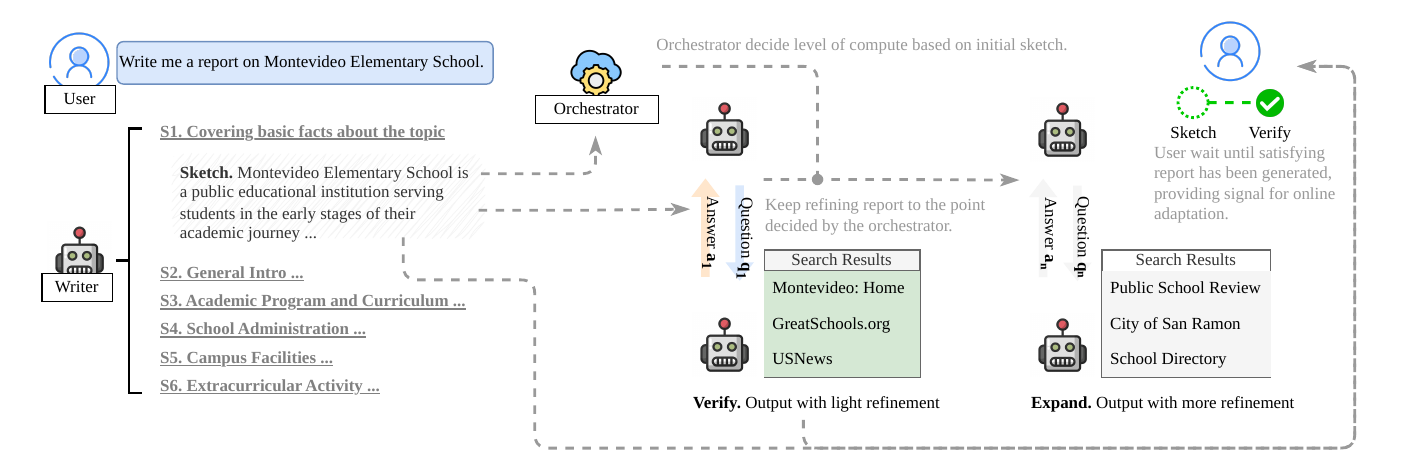}
    \caption{
      System overview of \frankenreport. User queries are decomposed into section-level plans (\textbf{gist}), with sections generated in parallel. Conditioned on early drafts, a trained action predictor enables adaptive stopping to optimize resource allocation. Partial outputs are displayed in real-time, allowing users to select preferred drafts for the final report while background refinements continue.}
    \label{fig:system-overview}
  \end{figure*}

  \section{\frankenreport}

  \label{sec:method}
  \frankenreport improves computation utility by devoting different levels of compute to each section of a report.
  We start by modifying the STORM pipeline to be a sequence of steps gradually \textit{refining} a coarsely drafted section gist 
  as shown in \autoref{fig:gantt-plot}; thus, drafts at different steps can be directly consumed by an end user if the quality is sufficient.
  An action predictor monitors these intermediate drafts and predicts whether further computation is likely to improve perceived quality, providing the hooks needed for confidence-aware early exits and personalized latency-quality trade-offs.

  Given a user query $x \in \mathcal{X}$, which is a knowledge-seeking request,
  the deep research system $\mathrm{P}$ produces
  a structured report $\mathbf{y}$ consisting of multiple sections $(y_1, y_2, \dots, y_n) \in \mathcal{Y}$. A \textit{configuration} fully specifies how a system should be applied to a given query.
  Different configurations of $\mathrm{P}$---for example, selecting which modules
  to run or how many reasoning steps to perform---lead to different outputs and computational costs. In this work, the configuration subspace $\mathcal{L}$ is described in \autoref{sec:step-wise-report-gen}.
  To adapt these choices dynamically, we introduce an \textit{orchestration policy}
  \[
    l : \mathcal{X} \times \mathcal{P} \mapsto \mathcal{L},
  \]
  which maps each query $x$ (and optionally the current model state $\mathrm{P}$)
  to a configuration $l(x, \mathrm{P}) \in \mathcal{L}$.
  Running system $\mathrm{P}$ under this configuration is represented by the operator
  \[
    \mathcal{T}(\mathrm{P}, x; l) = \mathrm{P}\big(x; l(x, \mathrm{P})\big).
  \]

 We define a \textit{quality function}
  $Q : \mathcal{X} \times \mathcal{Y} \mapsto \mathbb{R}$,
  which measures qualitative dimensions like the usefulness or factual quality of a generated report
  for a given query, and a \textit{cost function}
  $\mathrm{Cost} : (\mathrm{P}, x, l(x, \mathrm{P})) \mapsto \mathbb{R}_{\ge 0}$,
  which captures the time or compute required to run $\mathrm{P}$.
  In all experiments, $\mathrm{Cost}$ is a normalized, latency-calibrated stage cost rather than an API-token count: promoting a section from \textbf{Sketch} to \textbf{Verify} costs one unit and promoting it from \textbf{Sketch} to \textbf{Expand} costs roughly four. These weights follow the empirical mean transition-time ratio in our logs. On the rough wall-clock timeline in \autoref{fig:gantt-plot}, the zero-cost base corresponds to the approximately 16-second Sketch checkpoint, the one-unit promotion to the approximately 24-second Verify checkpoint, and the four-unit action to completing the full Expand path. The simulator sums these promotion units to impose a common allocation budget across sections.

  This distinction matters in deployment. Sections and later-stage work can overlap, and we launch the next stage speculatively while the lightweight regression-head predictor runs off the critical path, aborting that work if the predictor chooses to stop. Parallel and speculative execution can reduce user-perceived waiting time without actually reducing the aggregate work done by models; only refinements that are never launched or are terminated early reduce both latency and actual compute. 
  Accordingly, our curves measure quality under a latency-calibrated allocation proxy, not exact token, energy, or GPU-hour savings.

  The objective of \frankenreport is to learn an orchestration policy $l^\star$ achieving the best expected report quality while staying within an average computational budget $C$:

  \begin{equation}
    \begin{aligned}
      \max_{l : \mathcal{X} \times \mathcal{P} \to \mathcal{L}} \quad &
      \mathbb{E}_{x \sim \mathcal{X}}
      \!\left[
        Q\big(x, \mathcal{T}(\mathrm{P}, x; l)\big)
      \right] \\
      \text{s.t.} \quad &
      \mathbb{E}_{x \sim \mathcal{X}}
      \!\left[
        \mathrm{Cost}\big(\mathrm{P}, x, l(x, \mathrm{P})\big)
      \right]
      \le C.
    \end{aligned}
    \label{eq:constrained-optimization}
  \end{equation}

  This budgeted form directly matches the simulation in \autoref{fig:credit-assignments}. In a deployment where no hard cap is required, the same trade-off can instead be expressed as reward minus a user-specific cost penalty; \autoref{sec:learning-from-user-feedback} learns the quality side of that preference from online feedback.

  \subsection{Step-wise Report Generation}
  \label{sec:step-wise-report-gen}

  We factor $\mathcal{T}(\mathrm{P}, x; l)$ into three ordered stages.
  The report after each stage $s$ is denoted by $\mathbf{y}^{(s)}$.
  Choosing a configuration $l(x,\mathrm{P})$ then amounts to deciding which stages to execute, trading off the quality gain measured by $Q$ against the $\mathrm{Cost}$.

  For each query $x$, we first retrieve a document set $\mathcal{D}_x$ using only the user prompt.
  The retriever produces top-level section gists $\mathbf{g} = (g_1, \ldots, g_n)$ that summarize the intent of the final report $\mathbf{y}$ while remaining inexpensive because they avoid the multi-turn dialogue of the original STORM workflow, a modification we call \textit{gist-priming}. We allocate the $K$ evidence-gathering threads evenly across the gists so that the orchestrator can later decide, at section granularity, whether to continue or to stop. The resulting pipeline is illustrated in \autoref{fig:system-overview}, which exposes three early-exit options:

  \paragraph{Sketch} produces $\mathbf{y}^{(\text{sketch})}$ directly from the parametric knowledge of $\mathrm{P}$, conditioned on the gists $\mathbf{g}$. It minimizes latency but omits section-level evidence retrieval and grounding.

  \paragraph{Verify} extends to the fact-checking stage and yields $\mathbf{y}^{(\text{verify})}$. Each section performs a single retrieval round guided by its \textbf{Sketch} draft and gist $g_i$, mirroring outline-driven RAG without invoking the full dialogue planner.

  \paragraph{Expand} further runs the complete STORM workflow and outputs $\mathbf{y}^{(\text{expand})}$. Multi-turn dialogue retrieval proceeds independently for each section.

  \autoref{fig:gantt-plot} shows that with gist-priming, \textbf{Sketch} and \textbf{Verify} outcomes can be provided to a user much faster, even before the standard STORM outline can be formed. This allows more aggressive compute savings when the action predictor decides to early-exit.

  \subsection{Score-based Orchestration}
  \label{sec:score-based-moderator-training}

  We collect execution logs $\mathcal{S}$ consisting of tuples $(x, g_i, y^{(\text{sketch})}_i, y^{(\text{verify})}_i, y^{(\text{expand})}_i)$ and the corresponding costs for each section $i$. For $s,t \in \{\text{sketch},\text{verify},\text{expand}\}$ with $s$ preceding $t$, we estimate the marginal quality gained by advancing from $s$ to $t$,
  \[
    \Delta_i^{(s \rightarrow t)} = Q\big(x, y^{(t)}_i\big) - Q\big(x, y^{(s)}_i\big),
  \]
  and pair it with the cost difference
  \[\Delta \mathrm{Cost}^{(s \rightarrow t)} = \mathrm{Cost}_i(\mathrm{P}, x, t) - \mathrm{Cost}_i(\mathrm{P}, x, s).
  \]
  Here, $\mathrm{Cost}_i(\mathrm{P},x,s)$ is section $i$'s contribution to the total cost when it exits at stage $s$.
  We define the expected value of computation for advancing from $s$ to $t$ as the conditional expected marginal gain
  \[
    \mathrm{EVC}_i^{(s \rightarrow t)} = \mathbb{E}\!\left[\Delta_i^{(s \rightarrow t)} \mid x, g_i, y_i^{(s)}\right],
  \]
  which the policy considers together with the transition cost under the overall budget.

  An LLM with a regression head, denoted $f_\phi$, encodes the input tuple $(x, g_i, y^{(s)}_i)$ and estimates this conditional expected gain.
  Because later stages consume earlier drafts, the orchestration policy can encode $y^{(s)}_i$ in addition to the query $x$ without additional generation cost.

  \subsection{User Interface Design}
  \label{sec:interface-design}

  The user interface mirrors the staged decomposition of $\mathcal{T}(\mathrm{P}, x; l)$ and
  surfaces the evolving report $\mathbf{y}^{(s)}$ as soon as each stage finishes. \frankenreport comes with an intuitive UI to allow users to interact with and give feedback on these stage outputs, as shown in \autoref{fig:interface-overview} in \autoref{sec:interface-details}. A progress
  ribbon highlights the currently available drafts from $\{\text{sketch}, \text{verify}, \text{expand}\}$, while
  deeper reasoning continues in the background.
  This streaming view keeps users anchored in the
  report structure defined by the gists $\mathbf{g}$ without waiting for the full STORM
  dialogue to resolve.

  If for the initial draft $y^{(\text{sketch})}_i$, the estimated marginal gain
  $\hat{\Delta}_i^{(s \rightarrow t)} = f_\phi(x, g_i, y^{(s)}_i)$ leads the orchestration policy to decide that further computation is not worthwhile, later stages $y^{(\text{verify})}_i$ and $y^{(\text{expand})}_i$ are neither run nor shown. A fixed threshold can target a budget estimated from pilot runs; \autoref{sec:learning-from-user-feedback} instead adapts exit decisions online to user preferences.
  When the projected gain justifies additional compute, the interface prefetches stages further down the pipeline, further reducing the latency perceived.

  Once a draft from any early-exiting point (or full execution) is ready, the user can view that draft in a side panel as shown in \autoref{fig:dropdown-illustration}. The \textbf{Verify} and \textbf{Expand} generation is usually paired with citations, and the user can choose to replace the first draft from \textbf{Sketch} with any of them.

  \section{Experiment Results}

    We evaluate our adaptive exiting framework through three research questions: \textbf{RQ1}. \textit{Does increased compute (deeper polishing, richer retrieval) improve long-form quality?} \textbf{RQ2}. \textit{Can the orchestrator reliably predict these gains early?} \textbf{RQ3}. \textit{Which quality dimensions do users prioritize, and how can we adapt to them?} We address these below using datasets, ablations, predictive modeling, and human preference analysis.

  \subsection{Datasets}
  \label{sec:dataset}
  We draw topics and entities from five established long-form generation datasets to study staged generation and budget allocation over Sketch/Verify/Expand drafts. The collection emphasizes newer topics lacking complete Wikipedia coverage while retaining a subset representative of typical user queries: all 100 FreshWiki topics (the in-distribution source introduced with STORM) plus 900 prompts sampled from four larger auxiliary pools (1{,}000 prompts and 17{,}106 section-level examples in total; see \autoref{tab:section-stats}):

  \textbf{LongFact (LF)}~\citep{wei2024long} provides GPT-4-generated information-seeking queries that elicit multi-paragraph responses; we sample 100 prompts from its ``LongFact-Objects'' subset. \textbf{FActScore (FS)}~\citep{min-etal-2023-factscore} contains Wikipedia biographies at varied popularity levels, from which we sample 100 topics. \textbf{Core (CORE)}~\citep{jiang-etal-2025-core} targets factuality over a broader topic set that avoids overlap with FS; we sample 200 topics. \textbf{WildHallucinations (WH)}~\citep{zhao2024wildchat,zhao2024wildhallucinations} draws WildChat topics checked for Wikipedia coverage; we sample 500 entities \emph{without} Wikipedia pages. \textbf{FreshWiki (FW)}~\citep{shao2024assisting} is the STORM Wikipedia-article dataset; we use all 100 recently-edited, multi-section topics.
  \begin{table}[t]
    \centering
    {\small
      \begin{tabular}{@{}lcrrrrr@{}}
        \toprule
        &  & \textbf{CORE} & \textbf{FS} & \textbf{FW} & \textbf{LF} & \textbf{WH} \\
        \midrule
        \textbf{\#D} &  & 200 & 100 & 100 & 100 & 500 \\
        \textbf{\#S} &  & 3459 & 1761 & 1686 & 1749 & 8451 \\
        \textbf{S/D} &  & 17.3 & 17.6 & 16.9 & 17.5 & 16.9 \\
        \multirow{3}{*}{\textbf{T/S}} & \textbf{Ske} & 390.7 & 297.9 & 409.6 & 441.3 & 400.0 \\
        & \textbf{Ver} & 439.5 & 414.4 & 451.4 & 475.8 & 464.7 \\
        & \textbf{Exp} & 999.4 & 882.0 & 1026.7 & 1175.8 & 1110.1 \\
        \bottomrule
      \end{tabular}
    }
    \caption{Per-document (D), section (S) and token (T) statistics across drafts. A/B is ``number of A per B''.}
    \label{tab:section-stats}
  \end{table}
  
  For each dataset, we run our modified STORM pipeline with Llama-3.1-70B-Instruct \citep{grattafiori2024llama} to generate the three versions of outputs: Sketch (Ske), Verify (Ver), and Expand (Exp). We use Serper\footnote{
    \url{https://serper.dev/}
  } as our search backend.

  \subsection{Quality Assessment}
  \label{sec:quality-assessment-main}

  To determine the effect of additional compute on improving quality, we run five textual quality evaluation metrics $\mathrm{Q}$
  on the generation derived from different early exiting points in \autoref{tab:quality-metrics}. Later drafts with more compute tend to have higher quality.

  \autoref{tab:quality-metrics} reports \emph{aggregate} trends across sections. Mean gains vary by metric and are modest for Coherence and Engagingness. \autoref{fig:credit-assignments} evaluates how selectively allocating compute translates predicted gains into quality under fixed budgets. \textbf{Verify} also occasionally underperforms \textbf{Sketch} on style-oriented metrics: a single retrieval round can inject loosely relevant evidence and disrupt a fluent draft even while improving grounding, consistent with the ordering in Coherence ($4.39$ vs.\ $4.31$) but not Factuality ($.60$ vs.\ $.62$). Results with GPT-4.1 show that refinement margins also depend on the backbone's base capability and metric ceiling (\autoref{sec:gpt41-generalization}).

  \begin{table}[htbp]
    \centering
    \small
    \setlength{\tabcolsep}{2pt}
    \begin{tabular}{@{}lcccc@{}}
      \toprule
      \textbf{Metric} & \textbf{Range} & \textbf{Ske} & \textbf{Ver} & \textbf{Exp} \\
      \midrule
      \textbf{Coherence} & (1-5) & $4.39{\pm}.33$ & $4.31{\pm}.45$ & $\mathbf{4.48{\pm}.29}$ \\
      \textbf{Engagingness} & (1-3) & $2.05{\pm}.99$ & $2.07{\pm}.99$ & $\mathbf{2.12{\pm}.94}$ \\
      \textbf{Organization} & (1-5) & $4.66{\pm}.91$ & $4.69{\pm}.51$ & $\mathbf{4.90{\pm}.31}$ \\
      \textbf{Informativity} & ($>$ 0) & $.80{\pm}.26$ & $.84{\pm}.27$ & $\mathbf{.94{\pm}.26}$ \\
      \textbf{Factuality} & (0-1) & $.60{\pm}.21$ & $.62{\pm}.22$ & $\mathbf{.75{\pm}.14}$ \\
      \bottomrule
    \end{tabular}
    \caption{Mean $\pm$ standard deviation of quality metrics for $y^{s}$, where $s \in \{$Sketch, Verify, Expand$\}$, across sections.
    }
    \label{tab:quality-metrics}
  \end{table}

  \paragraph{LLM-as-a-Judge} We run multiple LLM-as-a-Judge metrics for textual qualities.
  These include \textit{Coherence}, for which we use the prompt from the \textbf{deepeval} framework.\footnote{\url{https://github.com/confident-ai/deepeval}}
  We also write our own prompts for Engagingness and Organization, following the criterion-based LLM-as-a-judge protocol of \citet{liu2023g} and the definitions used by \citet{shao2024assisting}, as detailed in \autoref{sec:evaluation-prompts}.

  \paragraph{Informativity} Following \citet{jiang-etal-2025-core}, we use Conditional Pointwise Mutual Information (CPMI) to evaluate the informativity of the text. For a claim $c$ decomposed from a section $y_i$,

  \begin{equation*}
    w_{\text{Info}}(c) = - \log \text{Pr}\big(c|\mathcal{H}(y_i)\big),
  \end{equation*}

  \noindent where $\mathcal{H}(y_i)$ is a set of trivial claims that can be assumed for the given section content. We use the model by \citet{wang2025always} to estimate the conditional probability of each claim happening \citep{chen-etal-2020-uncertain}. \footnote{\url{https://huggingface.co/Zhengping/conditional-probability-regression}}
  E.g., when generating claims for \textit{David Beckham}, $\mathcal{H}(y_i)$ may include claims like \textit{``David Beckham is a person.''} etc.
  While \citet{jiang-etal-2025-core} rely on manually written trivial claims, in our case since $y_i$ is generated from outline gist $g_i$ as described in \autoref{sec:method}, we decompose $g_i$ to get $\mathcal{H}(y_i)$. To avoid biases towards longer generation, the informativity of a section is calculated as

  \begin{equation*}
    Q_{\text{Info}}(x, y_i) = \frac{1}{|C_i|}\sum_{c \in C_i} w_{\text{Info}}(c),
  \end{equation*}

  Where $C_i$ is the set of claims from $y_i$.

  \paragraph{Factuality} We estimate factuality using SAFE \citep{wei2024long}, a search-based FActScore variant \citep{min-etal-2023-factscore}. To reduce the cost of large-scale web searches, we avoid full claim decomposition. While sentence-level search (e.g., VeriFastScore \citep{rajendhran2025verifastscore}) offers limited speedup in our setting, accurate estimation still requires claim-level search. Consistent with prior work \citep{krishna-etal-2023-longeval}, we find that sampling a small number of claims suffices (\autoref{sec:qual-additional-details-appendix}). We therefore randomly sample up to 10 claims per section to compute the factuality score.
  \subsection{Predicting quality differences}
    \begin{figure*}[!htpb]
    \centering
    \includegraphics[width=\textwidth]{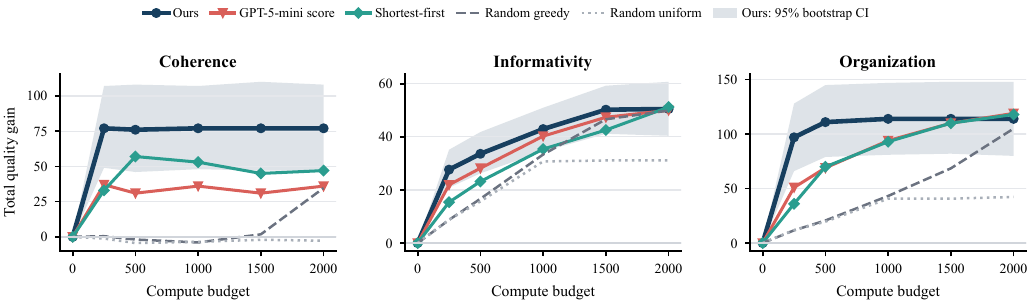}
    \caption{Total quality gain as the normalized compute budget increases. We compare our learned allocator with random-greedy, random-uniform, shortest-first, and zero-shot GPT-5-mini score routing. Shading is the 95\% nonparametric bootstrap interval for our deterministic policy (2{,}000 section-level resamples).}
    \label{fig:credit-assignments}
  \end{figure*}
  Following \S\ref{sec:score-based-moderator-training}, we finetune Llama-3.1-8B-Instruct \citep{grattafiori2024llama} with a regression head to predict the quality difference between drafts, using $l_2$ loss on score differences from execution logs across all datasets (up to 4 epochs, lr 5e-6, batch size 32, 8$\times$A100-80G; 8:1:1 train/dev/test split per dataset; best checkpoint by validation loss). For each quality assessment in \autoref{sec:quality-assessment-main}, we predict the gap using the user query (Q), the sketch (S), the verify draft (V), or both S and V as conditioning input, as shown in \autoref{tab:predictor-effectiveness}.
  
  Quality differences across drafts are reasonably predictable before generation. Organization and Coherence show the strongest correlations overall, while Informativity and Engagingness are also predictable for some transitions; query-only Factuality prediction is especially weak. This is expected: coherence and organization deficits are detectable without external knowledge, while factual gains depend on integrating evidence that is not yet available at the time of prediction. The gap underscores the importance of faithful uncertainty expression \citep{jiang2025conformal}, since hallucinations are harder to anticipate than abstention-induced gaps in structure or content.

  The \textbf{E}$\mid$\textbf{S} and \textbf{E}$\mid$\textbf{S},\textbf{V} rows predict different targets: the former estimates the total gain $\Delta^{(S \to E)} = Q(E) - Q(S)$ over the sketch, whereas the latter estimates the residual gain $\Delta^{(V \to E)} = Q(E) - Q(V)$ once $V$ is available. Because $V$ absorbs much of the predictable improvement over $S$, the residual target is smaller, rarer, and noisier, accounting for its lower correlations.

  \paragraph{Simulated Budget Allocation}
  We evaluate compute efficiency via a simulated allocation experiment. Given a fixed budget of $N$ additional normalized units across sections, we compare four operational baselines: (i)~\emph{random-uniform}, which samples an exit stage uniformly subject to feasibility; (ii)~\emph{random-greedy}, which randomly orders promotions and spends the remaining budget on the highest feasible stage; (iii)~\emph{shortest-first}, which promotes the currently shortest draft by one stage; and (iv)~\emph{GPT-5-mini score}, which zero-shot scores current drafts on the target quality dimension and promotes the lowest-scoring draft. The fixed always-Sketch/Expand policies are the two endpoints of the same action space; the upper endpoint corresponds to vanilla STORM~\citep{shao2024assisting} on every section. Per-layer, per-token, and per-answer early-exit methods in \autoref{sec:early-exit-related-work-anchor} use a different action space from this report-level allocation setting.
  As shown in \autoref{fig:credit-assignments}, conditioning allocation on the sketch output yields the strongest low-budget gains across all three dimensions, including against the shortest-first and zero-shot LLM routers. For dimensions that are cheap to improve (e.g., Coherence), allocating only 10\% of the budget recovers over 90\% of full-STORM performance; for dimensions that benefit from extended exploration (e.g., Informativity), gains rise smoothly with added compute. Relative to random allocation, gains reach up to $4\times$ in the low-budget regime. The remaining two dimensions and full baseline definitions are in \autoref{sec:credit-assignments-appendix}.

  \begin{table}[t]
    \centering
    \small
    \setlength{\tabcolsep}{4pt}
    \begin{tabular}{@{}l l ccc@{}}
      \toprule
      \textbf{Aspect} & \textbf{Target $\mid$ Conditioning} & \textbf{$r$} & \textbf{$\rho$} & \textbf{MSE} \\
      \midrule

      \multirow{6}{*}{Coherence}
      & V $\mid$ Q      & .353 & .170 & .670 \\
      & S $\mid$ Q      & .307 & .072 & .724 \\
      & V $\mid$ S      & \textbf{.746} & \textbf{.326} & .371 \\
      & E $\mid$ S      & .697 & .234 & .415 \\
      & E $\mid$ V      & .302 & .304 & \textbf{.339} \\
      & E $\mid$ S, V   & .071 & .070 & .509 \\

      \midrule

      \multirow{6}{*}{Engagingness}
      & V $\mid$ Q      & .088 & .090 & .227 \\
      & S $\mid$ Q      & .177 & .098 & .267 \\
      & V $\mid$ S      & \textbf{.587} & \textbf{.367} & \textbf{.098} \\
      & E $\mid$ S      & .493 & .318 & .374 \\
      & E $\mid$ V      & .283 & .281 & .131 \\
      & E $\mid$ S, V   & .237 & .229 & .123 \\

      \midrule

      \multirow{6}{*}{Informativity}
      & V $\mid$ Q      & .272 & .156 & .102 \\
      & S $\mid$ Q      & .366 & .254 & .048 \\
      & V $\mid$ S      & .535 & .315 & \textbf{.038} \\
      & E $\mid$ S      & \textbf{.554} & \textbf{.378} & .049 \\
      & E $\mid$ V      & .111 & .114 & \textbf{.038} \\
      & E $\mid$ S, V   & .111 & .115 & \textbf{.038} \\

      \midrule

      \multirow{6}{*}{Factuality}
      & V $\mid$ Q      & .003 & -.005 & .060 \\
      & S $\mid$ Q      & .000 & .016  & .072 \\
      & V $\mid$ S      & .149 & .080  & .081 \\
      & E $\mid$ S      & .281 & .254  & .070 \\
      & E $\mid$ V      & \textbf{.315} & \textbf{.291} & .102 \\
      & E $\mid$ S, V   & .289 & .252 & \textbf{.090} \\

      \midrule

      \multirow{6}{*}{Organization}
      & V $\mid$ Q      & .478 & .234 & .568 \\
      & S $\mid$ Q      & .273 & .244 & .635 \\
      & V $\mid$ S      & .821 & .418 & .231 \\
      & E $\mid$ S      & \textbf{.866} & .418 & \textbf{.155} \\
      & E $\mid$ V      & .486 & \textbf{.468} & .348 \\
      & E $\mid$ S, V   & .073 & .065 & .169 \\

      \bottomrule
    \end{tabular}
    \caption{Predictor performance across dimensions. In each target$\mid$conditioning label, the left side denotes the improvement target and the right side denotes the predictor input. \{\textbf{S}, \textbf{V}, \textbf{E}\} stands for \{Sketch, Verify, Expand\} drafts. \textbf{Q} stands for user query.}
    \label{tab:predictor-effectiveness}
  \end{table}
  \subsection{Online Learning from User Feedback}\label{sec:learning-from-user-feedback}
\begin{figure}[!htbp]
    \centering
    \includegraphics[width=\linewidth]{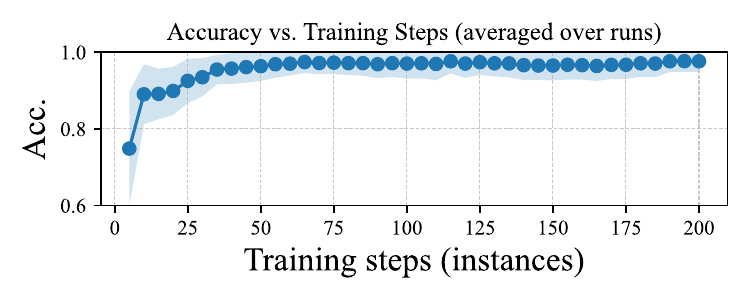}
    \caption{Online learning to adapt the orchestration policy to simulated user preference.}
    \label{fig:personalization}
  \end{figure}

  Using the same quality regression model, we can adapt to user preferences rather than fix the orchestration policy. Since users weight quality dimensions differently (e.g., some prefer brevity, others informativeness), we model each user as a preference vector $\theta$, where $\theta_i$ is the weight on dimension $i$. Running regression models for the metrics $\mathbf{Q}$ gives each draft $y_i^{(k)}$ a feature vector $\mathbf{q}_i^{(k)}$, and observing multiple drafts of a section $y_i^{(1)}, \dots, y_i^{(K)}$, the user's preference follows the ordering of the inner product
  \begin{equation}
    \label{eq:pref}
    k^* = \argmax_{k} \theta^T\mathbf{q}_i^{(k)} + \epsilon.
  \end{equation}
  which translates to $K - 1$ preference observations
  \begin{equation*}
    k^* \succ k,\quad \forall k \in [K] \setminus \{k^*\}.
  \end{equation*}
To test how effective the policy is at adapting to a user preference vector, we randomly sample 10 simulated user preferences $\theta \sim \text{Dirichlet}(1)$. We then fit a Bradley-Terry model to estimate $\hat{\theta}$ with independent Gaussian prior. In each round, we now sample an action based on the current posterior $\hat\theta$ to simulate one round of interaction with the user through our interface  as discussed in \autoref{sec:interface-design}. The user then provides their feedback according to \autoref{eq:pref}.
  \autoref{fig:personalization} plots policy accuracy in predicting the user-preferred exit point against training steps: accuracy improves steadily and reaches a high level within 10 examples. Since a user can only see drafts before the predicted exit point (e.g., predicting \textbf{Verify} hides \textbf{Expand}), we add an ``information weighting'' term to encourage early exploration; \autoref{sec:personalization} compares different information weights.
  
  \subsection{User Study}\label{sec:user-study}
\begin{figure}[htbp]
    \centering
    \includegraphics[width=\linewidth]{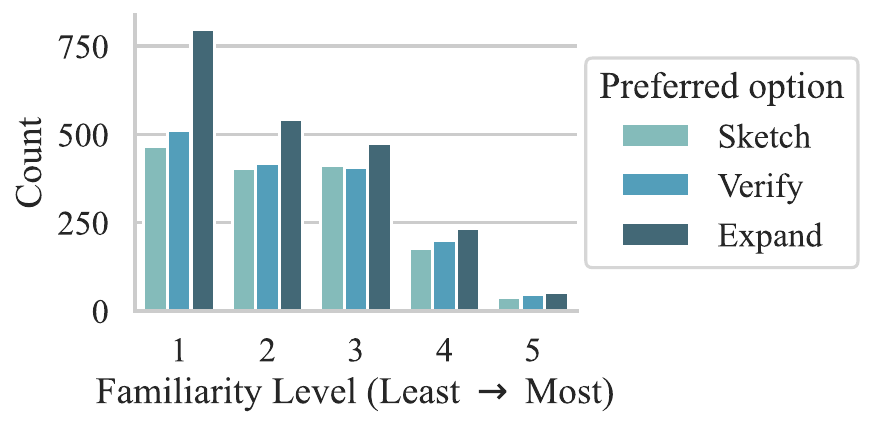}
    \caption{Preferences change with topic familiarity.}
    \label{fig:metric-correlation}
  \end{figure}

  We conduct a user study collecting human quality assessments on the different early-exit drafts, reusing the 100 test prompts and generations from \autoref{sec:dataset}. For each prompt we show a randomly shuffled set of the three generations $\{\text{Sketch}, \text{Verify}, \text{Expand}\}$ per section and ask for a 1--5 overall-quality rating; across 100 participants, each annotation task covers 5 topic reports, and we also collect self-reported confidence and topic familiarity along with reasons for their highest/lowest rankings. Citations are removed from all drafts for fair comparison.\footnote{As discussed in the introduction, users' subjective quality judgments could be biased by surface-level cues.} See \autoref{appendix:user-study} for details.
  Agreement among ratings was mild (Krippendorff's $\alpha = 0.43$), improving slightly with higher average familiarity or confidence (\autoref{appendix:user-study}). \autoref{fig:metric-correlation} shows how often each draft received the maximum score across familiarity levels. Later drafts are preferred more often overall, with \textbf{Expand} receiving the maximum score most frequently at every familiarity level. \textbf{Sketch} nevertheless remains preferred for a nontrivial subset of sections, motivating section-level early exiting.

  \paragraph{Automated metrics vs.\ human ratings} We measure alignment between the automated dimensions $\mathbf{Q}$ and human judgments by fitting a linear combination of the five scores to predict aggregated human ratings on the same generations, yielding in-sample $\mathrm{MSE} = 0.211$, Pearson $r = 0.582$, and Spearman $\rho = 0.384$. Automated metrics thus explain a meaningful but incomplete portion of perceived quality, supporting their use as a training signal alongside online personalization (\autoref{sec:learning-from-user-feedback}) for residual user-specific variation.

  \paragraph{Human-preference replay} We also replay the observed user sequence with a user-specific quality weighting initialized uniformly and updated only from that user's previously annotated topics. This online, out-of-sample procedure improves the selected human rating by $.042$ on the 1--5 scale over a fixed uniform weighting. Fitting all of a user's annotations and back-assigning selections---an in-sample counterpart approximating stable preferences after several interactions---improves it by $.119$. Most participants annotated at most five topics, placing the online result in a low-data personalization setting.
  \begin{table}[!htbp]
    \begin{center}
      \begin{tabular}{ccc}
        \toprule
        & \textbf{Acc} & \textbf{Acc (binary)} \\
        \midrule
        Majority & .406 & .716 \\
        GPT-5-mini & .649 & .811 \\
        +familiarity & .640 & .815 \\
        GPT-5.1 & .676 & .820 \\
        \bottomrule
      \end{tabular}
    \end{center}
    \caption{Profiling and user preference prediction results. We evaluate draft selection accuracy and ``Needs Improvement'' prediction (binary).}\label{tab:profiling}
  \end{table}

  \paragraph{Inferring User Preferences}
  We further study whether preferences can be inferred from written reasons. For each user we use up to 10 annotated reports, hold out one for validation, and use an LLM to summarize the user's preferences from the rankings and explanations of the rest, which have been demonstrated to provide useful feedback for preference learning \citep{jiang2026configurable}; this profile then predicts the preferred draft for each validation section once all drafts are available. \autoref{tab:profiling} shows that real-world preferences are identifiable: an LLM conditioned on a user's preference summary predicts future preferences far more accurately than a majority baseline.
  \autoref{fig:personalization} further shows the \frankenreport orchestrator reaches similar accuracy within 5 interaction rounds before any draft is generated, indicating efficient adaptation.

  \section{Conclusion}

We present \frankenreport, an interface to a deep research pipeline that exposes intermediate drafts to reduce compute and latency by reliably estimating final draft quality from early signals. Under fixed costs, \frankenreport consistently outperforms random expansion baselines, and although real-world user preferences vary across topics while generally favoring later drafts, our adaptive orchestrator learns them online with modest budgets. Future work includes topic-conditioned features and extending the Sketch/Verify/Expand instantiation to variable-depth refinement under the same expected-gain accounting.

\section*{Limitations}

We acknowledge several limitations in our work. First, our experiments focus exclusively on English-language report generation benchmarks, which does not guarantee generalization to other languages. Second, the magnitude and shape of staged quality gains depend on a model's base capability and metric ceilings. Our full orchestration study uses Llama-3.1-70B-Instruct; the GPT-4.1 study in \autoref{sec:gpt41-generalization} shows that such gains persist outside the Llama family, but covers only non-search-backed metrics.
Third, we model individual user preferences as a linear weighting over the five quality dimensions $\mathbf{Q}$; real user preferences can be nonlinear and context-dependent, and extending the formulation to richer (e.g., feature-interaction or contextual-bandit) preference models is a natural next step. Finally, we constrain our evaluation to reliably automatable metrics; future work incorporating heterogeneous or human-centric metrics may better capture real-world user preferences.

  \bibliography{custom}

  \appendix

  \section{Interface Design Details}
  \label{sec:interface-details}
\begin{figure*}[t]
    \centering
    \includegraphics[width=\linewidth,trim={.6cm .5cm .6cm .5cm},clip]{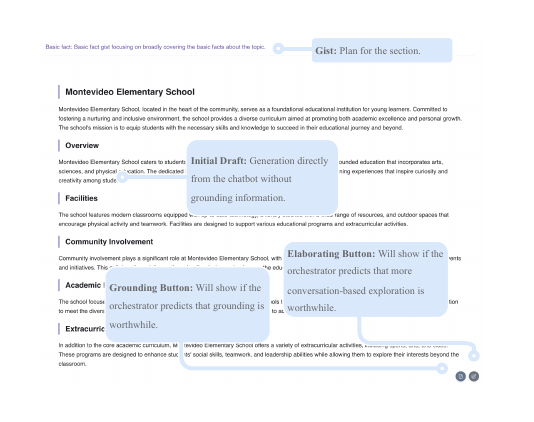}
    \caption{Overview of the \frankenreport interface that surfaces key controls and status indicators to the user. The initial draft uses a retrieval-informed gist but omits section-level evidence retrieval.}
    \label{fig:interface-overview}
  \end{figure*}

\begin{figure*}[t]
    \centering
    \includegraphics[width=\linewidth,trim={.5cm 1.0cm .5cm 1.0cm},clip]{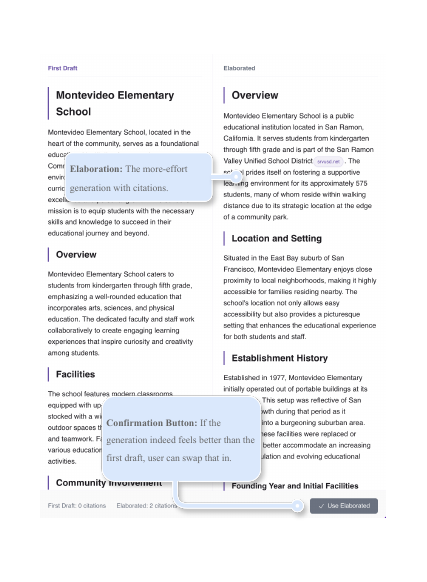}
    \caption{Draft-selection dropdown in the side panel, used to inspect alternative generations before confirming an exit decision.}
    \label{fig:dropdown-illustration}
  \end{figure*}

This appendix expands the interface description from Section~\ref{sec:interface-design}, breaking down the user interface into its major panels and controls. Figure~\ref{fig:interface-overview} labels the gist, initial draft, grounding control, and elaboration control. In the interface terminology, Initial Draft and First Draft correspond to \textbf{Sketch}, Grounding corresponds to \textbf{Verify}, and Elaborating or Elaborated corresponds to \textbf{Expand}. Figure~\ref{fig:dropdown-illustration} zooms in on the override menu used to inspect alternative drafts and provide corrective feedback before confirming an exit decision. Together, these illustrations provide the finer-grained component mapping referenced in the main text.

  \section{Evaluation Prompts}
  \label{sec:evaluation-prompts}

  \begin{figure*}[t]
    \centering
    \begin{minipage}{0.95\linewidth}
      \begin{llmjudgeprompt}[label={prompt:coherence-judge}]{Prompt: Coherence evaluation}
        \promptmeta{Model: \texttt{gpt-4.1}}

        \userrole\ Evaluate the coherence of the passage. Coherence is the collective
        quality of all sentences. The passage should be well-structured and
        well-organized. The passage should not just be a heap of related information,
        but should build from sentence to sentence to a coherent body of information
        about a topic and the gist of the section.\\
        \\
        Below is a rubric to help you evaluate the coherence of the passage:\\
        - 1 point: Poor coherence - The passage is disorganized, sentences don't connect logically, and information appears random or scattered.\\
        - 2 points: Below average coherence - Some logical flow exists but there are noticeable gaps in organization and connection between ideas.\\
        - 3 points: Average coherence - The passage has reasonable structure with most sentences connecting logically, though some improvements could be made.\\
        - 4 points: Good coherence - Well-organized passage with clear logical flow and strong connections between sentences and ideas.\\
        - 5 points: Excellent coherence - Exceptionally well-structured passage that builds seamlessly from sentence to sentence into a coherent body of information.\\
        \\
        After evaluating the passage, assign a coherence score from 1 to 5 based on the rubric above. Provide a brief explanation for your score, highlighting specific aspects of the passage that influenced your evaluation.
      \end{llmjudgeprompt}
    \end{minipage}
    \caption{Prompt used for coherence scoring.}
    \label{fig:prompt-coherence-judge}
  \end{figure*}

  \begin{figure*}[t]
    \centering
    \begin{minipage}{0.95\linewidth}
      \begin{llmjudgeprompt}[label={prompt:organization-judge}]{Prompt: Organization evaluation}
        \promptmeta{Model: \texttt{gpt-4.1}}

        \userrole\ You are an expert encyclopedia writer. Your task is to evaluate the
        quality of a Wikipedia-style article section based on whether it is well
        organized and logically structured. Based on the provided section, you should
        give a Likert-scale rating from 1 to 5, where 1 indicates very poor
        organization and 5 indicates excellent organization. Use the following
        rubrics and pair each numeric score with its label:\\
        \\
        1. \textbf{Very Poor Organization}: Disorganized; lacks logical structure and coherence.\\
        2. \textbf{Poor Organization}: A basic structure is present but inconsistently followed; frequent jumps or gaps.\\
        3. \textbf{Fair Organization}: Organized; a clear structure is mostly followed with some lapses in coherence.\\
        4. \textbf{Good Organization}: Well organized; clear sections and logical flow throughout with minor issues.\\
        5. \textbf{Excellent Organization}: Exceptionally well structured; coherent hierarchy and seamless transitions throughout.\\
        \\
        Respond with both the numeric score (1–5) and the paired label (e.g.,
        "3 -- Fair Organization"), plus a one-sentence rationale.\\
        \\
        Return your response in the following format:\\
        \\
        \texttt{<numeric score> -- <label>}\\
        \texttt{Rationale: <your rationale>}\\
        \\
        Here is the section to evaluate: \{section\}
      \end{llmjudgeprompt}
    \end{minipage}
    \caption{Prompt used for organization scoring.}
    \label{fig:prompt-organization-judge}
  \end{figure*}

  \begin{figure*}[t]
    \centering
    \begin{minipage}{0.95\linewidth}
      \begin{llmjudgeprompt}[label={prompt:engagingness-judge}]{Prompt: Engagingness evaluation}
        \promptmeta{Model: \texttt{gpt-4.1}}

        \userrole\ You are an expert editor evaluating how engaging an encyclopedia
        section feels to a curious reader. Judge whether the writing is captivating,
        offers valuable insights, and keeps the audience interested.\\
        \\
        Provide a score from 1 to 3 using the following rubric:\\
        \\
        1. Dull -- Generic, repetitive, or fails to provide interesting insights.\\
        2. Neutral -- Somewhat interesting but lacks depth or notable takeaways.\\
        3. Interesting -- Captivating, insightful, and keeps the reader engaged.\\
        \\
        Reply with the numeric score, followed by a short rationale in this exact format:\\
        \\
        \texttt{<numeric score>}\\
        \texttt{Rationale: <your rationale>}\\
        \\
        Topic: \{topic\}\\
        Gist: \{gist\}\\
        \\
        Section to evaluate:\\
        \{section\}
      \end{llmjudgeprompt}
    \end{minipage}
    \caption{Prompt used for engagingness scoring.}
    \label{fig:prompt-engagingness-judge}
  \end{figure*}

  \begin{figure*}[t]
    \centering
    \begin{minipage}{0.95\linewidth}
      \begin{llmjudgeprompt}[label={prompt:profiling}]{Prompt: User preference summarization}
        \promptmeta{Model: \texttt{gpt-4.1}}

        \userrole\ You are profiling one user's tastes from their evaluations of
        multiple sections in one or more reports. In each section the user read three
        draft alternatives and picked which draft was best and which was worst, then
        explained why. Each report line also carries the user's self-reported
        familiarity with that topic.\\
        Reports (JSON Lines; each line is a JSON object with keys
          \texttt{report\_title}, \texttt{familiarity}, and \texttt{sections\_jsonl}
          where \texttt{sections\_jsonl} is JSON Lines of section objects with keys
          \texttt{section\_number}, \texttt{section\_gist}, \texttt{options},
          \texttt{best\_index} (list or null), \texttt{worst\_index} (list or null),
        \texttt{best\_reason}, \texttt{worst\_reason}):\\
        \{reports\}\\
        \\
        Infer what the user consistently values and dislikes across all reports. Factor in how preference signals might shift with higher or lower familiarity, but keep the profile concise.\\
        \\
        Write 4--6 sentences in second person (``You …'') describing the user\'s preferences. Then add a heading `Preference Signals' followed by 2--4 hyphen bullets capturing the strongest likes/dislikes.
      \end{llmjudgeprompt}
    \end{minipage}
    \caption{Prompt used for user preference summarization.}
    \label{fig:prompt-profiling}
  \end{figure*}

  \begin{figure*}[t]
    \centering
    \begin{minipage}{0.95\linewidth}
      \begin{llmjudgeprompt}[label={prompt:cls-selection}]{Prompt: Preference-conditioned option selection}
        \promptmeta{Model: \texttt{gpt-4.1}}

        \userrole\ You are predicting which option this user will rate highest.\\
        User profile:\\
        \{profile\}\\
        \\
        Report: \{report.title\}\\
        User familiarity with this topic: \{report.familiarity or 'Unknown'\}\\
        Section \{section.number\}: \{section.gist or 'N/A'\}\\
        Options:\\
        \{*options\_lines\}\\
        \\
        Respond with a single integer 1-\{len(section.options)\} for the option most aligned with the user's preferences.
      \end{llmjudgeprompt}
    \end{minipage}
    \caption{Prompt used for preference-conditioned option selection.}
    \label{fig:prompt-cls-selection}
  \end{figure*}

  Together, these prompts span key evaluation dimensions: coherence scoring checks whether a passage logically builds from sentence to sentence (\autoref{fig:prompt-coherence-judge}); organization scoring verifies structural clarity in Wikipedia-style sections (\autoref{fig:prompt-organization-judge}); engagingness scoring judges how captivating and insightful a section feels to curious readers (\autoref{fig:prompt-engagingness-judge}); preference summarization distills a user's likes and dislikes across reports to guide orchestration (\autoref{fig:prompt-profiling}); and preference-conditioned option selection predicts which draft a user will most prefer given their profile and past choices (\autoref{fig:prompt-cls-selection}).

\section{Generalization to a Second Chat Model}
\label{sec:gpt41-generalization}

Our full orchestration experiments in \autoref{sec:quality-assessment-main} use Llama-3.1-70B-Instruct and couple report generation with search-backed evaluation across thousands of topics. We complement them with GPT-4.1 results on all non-search-backed metrics, excluding Factuality, which depends on retrieval and evidence integration. \autoref{tab:gpt41-quality-metrics} summarizes these results.

\begin{table}[htbp]
  \centering
  \begin{tabular}{@{}lccc@{}}
    \toprule
    \textbf{Metric} & \textbf{Ske} & \textbf{Ver} & \textbf{Exp} \\
    \midrule
    \textbf{Coherence} (1--5) & 4.930 & \textbf{4.982} & 4.739 \\
    \textbf{Engagingness} (1--3) & 2.888 & 2.931 & \textbf{2.948} \\
    \textbf{Organization} (1--5) & 4.889 & \textbf{4.993} & 4.986 \\
    \textbf{Informativity} ($>$ 0) & .831 & .871 & \textbf{1.020} \\
    \bottomrule
  \end{tabular}
  \caption{Quality metrics with GPT-4.1 as the generation model, averaged across all test queries. Factuality is omitted because it requires search-backed evaluation.}
  \label{tab:gpt41-quality-metrics}
\end{table}

Qualitatively, these results show that measurable quality gains persist for some metrics but not uniformly across later passes, consistent with the Llama-based results in \autoref{tab:quality-metrics}. A notable difference is that GPT-4.1 produces substantially higher first-pass quality: for instance, Coherence starts near the ceiling of the 1--5 scale ($4.93$ vs.\ $4.39$ for Llama). As a consequence, the margins among computationally lightweight dimensions---those that are relatively easy to improve with additional passes, such as Coherence and Organization (\autoref{fig:credit-assignments})---tend to shrink, and in some cases the \textbf{Verify} draft scores above the \textbf{Expand} draft, most notably for Coherence ($4.982$ vs.\ $4.739$). In contrast, Informativity gains remain substantial ($0.831 \to 1.020$), mirroring the trend observed with Llama and confirming that content enrichment is the dimension that benefits most from deeper exploration.

These differences change the shape of metric-specific quality--cost trade-off curves: with a stronger base model, the orchestrator can skip refinement on dimensions already near saturation and concentrate compute on dimensions with greater room for improvement. Future-utility estimation therefore remains useful for directing refinement toward dimensions with available headroom.

  \section{Budget Allocation}
  \label{sec:credit-assignments-appendix}

\begin{figure}
    \centering
    \includegraphics[width=\linewidth]{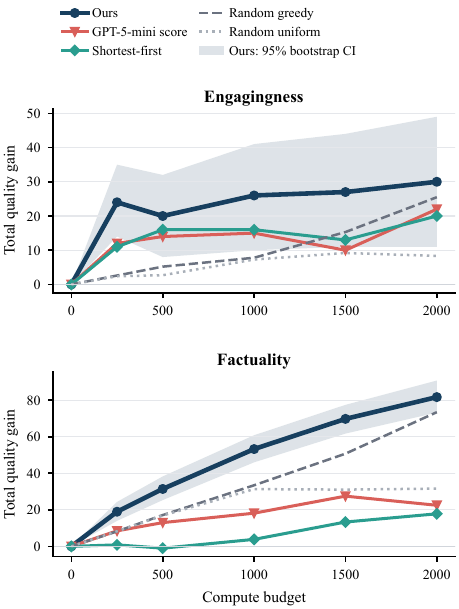}
    \caption{Budget allocation for engagingness and factuality with the same four baselines as \autoref{fig:credit-assignments}. Shading is the 95\% section-level bootstrap interval for our deterministic policy.}
    \label{fig:credit-assignments-fact-eng}
  \end{figure}

  \autoref{fig:credit-assignments-fact-eng} reports the corresponding budget-allocation results for \textbf{Factuality} and \textbf{Engagingness}.

  The baseline configurations are defined below.

  \paragraph{Random-uniform} This baseline samples an action for each section independently and uniformly at random from the set $\{\textbf{Sketch},\textbf{Verify},\textbf{Expand}\}$, without conditioning on any section-level features. Sampling continues until a feasible allocation that satisfies the total budget constraint is obtained, at which point the process terminates.

  \paragraph{Random-greedy} This baseline randomly orders the sections and assigns the highest-effort action permitted by the remaining budget to each section in turn, prioritizing \textbf{Expand} over \textbf{Verify} and \textbf{Verify} over \textbf{Sketch}. It selects \textbf{Expand} if the budget allows; otherwise it falls back to \textbf{Verify}, and finally to \textbf{Sketch}. Because \autoref{tab:quality-metrics} shows that later stages improve average quality, the strategy concentrates compute on the highest feasible stage.

  \paragraph{Shortest-first} At every step this heuristic promotes the currently shortest section draft by one stage (\textbf{Sketch}$\rightarrow$\textbf{Verify}$\rightarrow$\textbf{Expand}), subject to the remaining budget. It tests a simple operational proxy for underdeveloped content without using learned quality predictions.

  \paragraph{GPT-5-mini score} We zero-shot prompt GPT-5-mini to score each current draft on the target quality dimension and promote the lowest-scoring feasible draft. Scores are recomputed from the stage currently visible to the policy. This is a draft-aware LLM router, but unlike our allocator it is not trained to predict the \emph{marginal gain} of a particular next stage.

  \section{Additional Details on Quality Assessment}
  \label{sec:qual-additional-details-appendix}
  \begin{figure*}[htbp]
    \centering
    \begin{minipage}{0.48\linewidth}
      \centering
      \includegraphics[width=\linewidth]{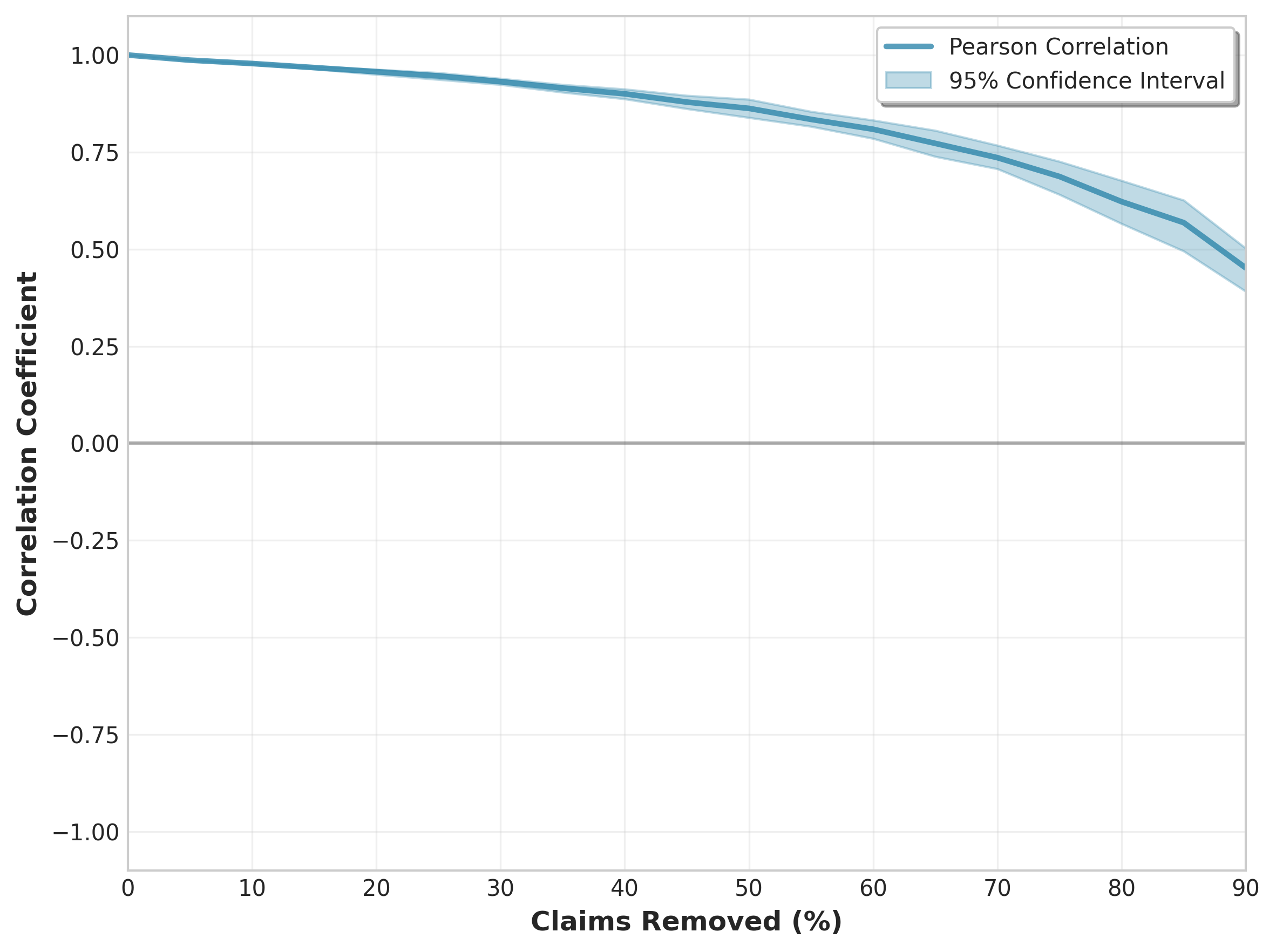}
    \end{minipage}\hfill
    \begin{minipage}{0.48\linewidth}
      \centering
      \includegraphics[width=\linewidth]{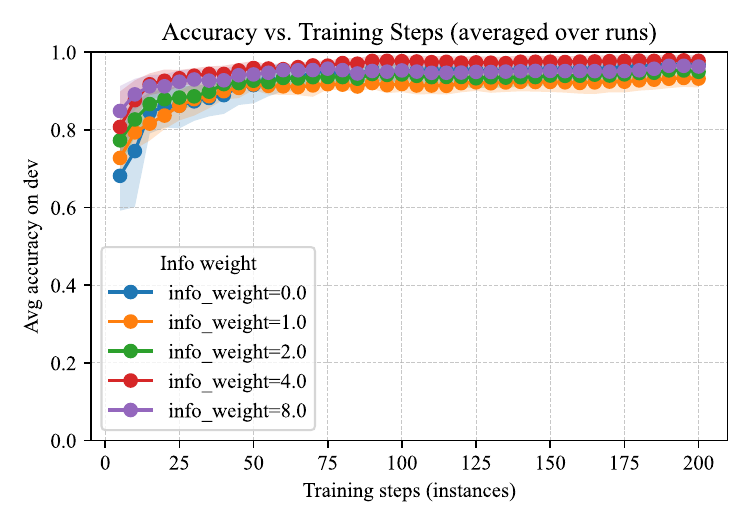}
    \end{minipage}
    \caption{Sample-efficiency results for factuality scoring (left) and user-preference adaptation under different information weights $\alpha$ (right).}
    \label{fig:sample-efficiency-and-personalization}
  \end{figure*}
  \paragraph{Factuality} \autoref{fig:sample-efficiency-and-personalization} (left) shows that factuality scores remain stable when claims are subsampled. We therefore evaluate a subset of claims as specified in \autoref{sec:quality-assessment-main}.
  \paragraph{License} The \textbf{LongFact} prompts \citep{wei2024long} are released under CC BY 4.0, while the accompanying software is under Apache 2.0. \textbf{FActScore} \citep{min-etal-2023-factscore} and \textbf{WildHallucinations} \citep{zhao2024wildhallucinations} are released under the MIT License. \textbf{FreshWiki} \citep{shao2024assisting}, which is derived from Wikipedia, is released under CC BY-SA. The public \textbf{Core} repository \citep{jiang-etal-2025-core} did not specify a license at the time of access. The Llama 3.1 herd of models \citep{grattafiori2024llama} is released under the Llama 3.1 Community License.

  \section{User Preference Adaptation}
  \label{sec:personalization}
  To balance exploration and exploitation during preference adaptation, we add an information-weight term to action selection. This improves early sample efficiency at the cost of locally suboptimal predictions under the current orchestration policy. At each step, action $k$ is selected according to the score function parameterized by $\hat{\theta}$ and its expected information gain:
  \begin{equation*}
    \hat{k}^* = \argmax_{k} \left(\hat{\theta}^T\mathbf{q}_i^{(k)} + \alpha w_{\text{Info}}^{(k)}\right).
  \end{equation*}
  This biases action selection toward running the full pipeline (\textbf{Expand}) at the beginning of the adaptation process,
  thus collecting more informative feedback from the user. \autoref{fig:sample-efficiency-and-personalization} (right) shows the adaptation results with different $\alpha$ values.

  Larger $\alpha$ values lead to faster adaptation but have little effect on final accuracy, indicating that the default policy already explores sufficiently for efficient preference adaptation.
  \section{User Study Details}
  \label{appendix:user-study}

  \subsection{Annotation}
  We conduct a user study to gather human quality assessments of the generations produced at different early-exit points. This study was reviewed and approved by an internal Institutional Review Board.
  We use the same 100 test prompts and generations from the full dataset described in \autoref{sec:dataset}.
  In each annotation task, a participant was assigned to evaluate generated content for 5 different test prompts (topics), with 100 participants in total. Participants could complete more than one annotation task.
  Therefore, we obtained 5 independent evaluations per test prompt.

  We recruited 100 participants from Prolific.
  We used Prolific's standard sampling distribution.
  Using Prolific's screening feature, we screened for primarily English-speaking participants
  who were located in the United States, had a 100\% approval rate,
  had submitted at least 20 prior AI-evaluation tasks,
  and held an undergraduate or higher degree.

  Each annotation task consisted of one Qualtrics survey.
  The survey asked participants to pretend that they are tasked to generate an informational report on a topic.
  Each survey included five topic reports they were asked to review and evaluate.
  The survey began with a consent form providing detailed information about the study.
  Once the participant provided consent, the survey guided them through each of the five topics.

\begin{figure*}[!htbp]
  \includegraphics[width=\textwidth]{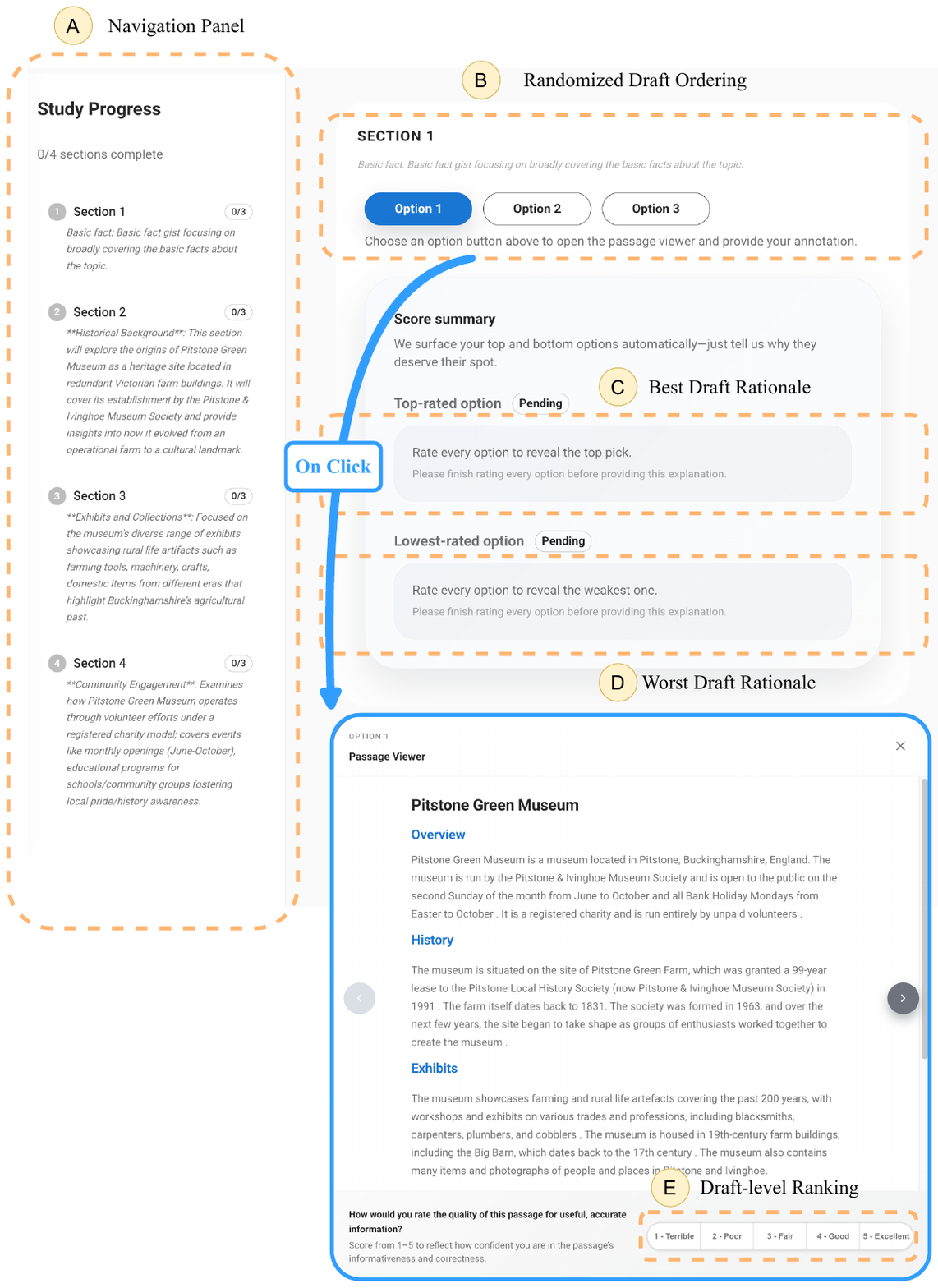}
  \caption{Screenshot of the user study interface for ``Pitstone Green Museum'' topic.}
  \label{fig:study-interface}
\end{figure*}

  For each topic, they were asked to navigate to a webpage on a separate browser window (Figure~\ref{fig:study-interface}).
  The webpage presented an AI-generated report containing sections.
Each section was presented with a randomly shuffled set of all three generations $\{\text{Sketch}, \text{Verify}, \text{Expand}\}$ (Figure~\ref{fig:study-interface}B). For each generation option, participants were asked to rate its quality (Figure~\ref{fig:study-interface}E). Once they had evaluated all sections and options, they copied the resulting JSON text into the corresponding section of the survey. They were then asked two questions: (1)~How familiar are you with the topic of this report? (1 = Not familiar at all, 5 = Extremely familiar); and (2)~How confident are you in your assessment of the report's quality? (1 = Not confident at all, 5 = Extremely confident).
The survey asked participants to repeat this process for each of the five topics.

After reviewing and evaluating all five topic reports, participants answered demographic questions covering age, gender identity, race/ethnicity, and education level. In total, the survey took approximately 90 minutes, and each participant who completed it was compensated USD \$19.

\subsection{Inter-Annotator Agreement Analysis}
\label{sec:iaa-analysis}

\begin{figure*}[!htbp]
  \centering
  \begin{minipage}{0.49\textwidth}
    \centering
    \includegraphics[width=\linewidth]{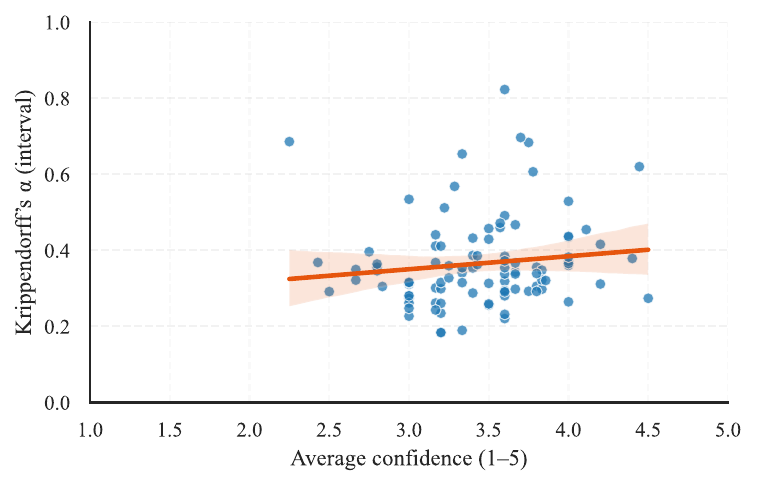}
  \end{minipage}\hfill
  \begin{minipage}{0.49\textwidth}
    \centering
    \includegraphics[width=\linewidth]{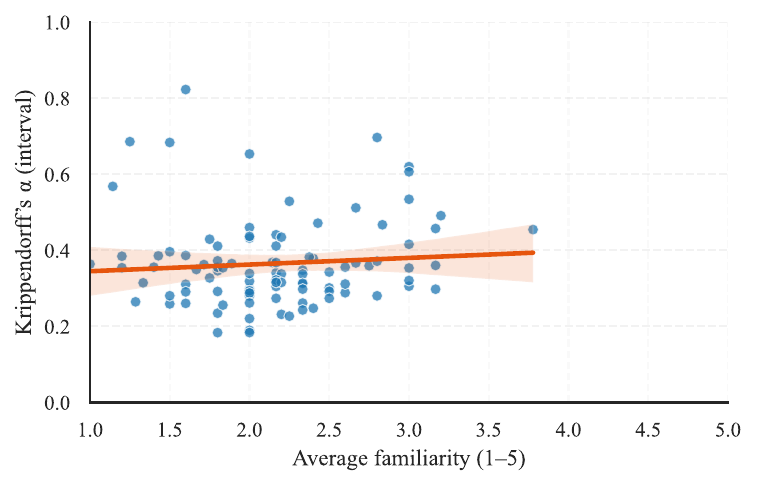}
  \end{minipage}
  \caption{Average inter-annotator agreement as a function of annotator-reported confidence (left) and topic familiarity (right).}
  \label{fig:oracle-pred-confidence}
\end{figure*}

We calculate inter-annotator agreement on section-draft-level ratings using Krippendorff's $\alpha$. Agreement among ratings is mild, with Krippendorff's $\alpha = 0.43$ (\autoref{sec:user-study}).
We further analyze how average section-draft-level Krippendorff's $\alpha$ varies with user-reported confidence and topic familiarity. \autoref{fig:oracle-pred-confidence} shows slightly higher agreement at higher average familiarity or confidence, although the effect is modest.

\subsection{Oracle Preference Modeling}
\label{sec:oracle-preference-modeling}
The oracle preference-modeling procedure underlying \autoref{tab:profiling} has access to more information than the online preference model of \autoref{sec:method}, providing an upper bound on attainable preference-prediction performance.

For each user, we adopt a leave-one-report-out protocol: we hold out one annotated report for validation and use the remaining reports to construct the user's preference \emph{profile} with the prompt in \autoref{fig:prompt-profiling}. The prompt conditions on the user's per-section best/worst draft choices, their free-text justifications, and their self-reported topic familiarity, and emits a concise second-person description paired with a short list of \emph{preference signals}.

Given this profile, we predict the user's preferred draft for each section of the held-out report using the preference-conditioned selection prompt in \autoref{fig:prompt-cls-selection}. This step sees \emph{all three} drafts $\{\text{Sketch}, \text{Verify}, \text{Expand}\}$ of every section at once, whereas the online orchestrator of \autoref{sec:method} observes only the drafts up to its predicted exit point and receives implicit, single-session feedback rather than explicit rationales. Access to all drafts and explicit rationales makes this procedure an \emph{oracle} that upper-bounds deployed preference-prediction accuracy.

We report two metrics in \autoref{tab:profiling}: draft-selection accuracy (which of the three drafts the user rates highest) and a binary ``Needs Improvement'' accuracy (whether the user prefers any further refinement over the initial \textbf{Sketch} draft). A majority-vote baseline that ignores the profile attains $.406$ selection and $.716$ binary accuracy. Conditioning on the LLM-summarized profile improves both substantially: GPT-5-mini reaches $.649/.811$ and GPT-5.1 reaches $.676/.820$; additionally surfacing the user's self-reported familiarity (\texttt{+familiarity}) leaves selection essentially unchanged ($.640$) while slightly improving the binary task ($.815$). These results show that, when the actual drafts and a rich preference profile are available, real-world user preferences are identifiable well above chance. Using only sequentially observed drafts and implicit feedback, the online orchestrator approaches this oracle accuracy within roughly five rounds of interaction (\autoref{fig:personalization}).

\end{document}